# Aerodynamic Force and Flow Structures of Two Airfoils in Flapping Motions

Shi Long Lan and Mao Sun

Institute of Fluid Mechanics, Beijing University of Aeronautics & Astronautics

Beijing 100083, P.R. China

Aerodynamic force and flow structures of two airfoils in tandem configuration performing flapping motions are studied, using the method of solving the Navier-Stokes equations in moving overset grids. Three typical phase differences between the fore- and aft-airfoil flapping cycles are considered. The following has been shown. (1) In the case of no interaction (single airfoil), the time average of the vertical force coefficient over the downstroke is 2.74, which is about 3 times as large as the maximum steady-state lift coefficient of a dragonfly wing; the time average of the horizontal force coefficient is 1.97, which is also large. The reasons for the large force coefficients are the acceleration at the beginning of a stroke, the delayed stall and the "pitching-up" motion near the end of the stroke. (2) In the cases of two-airfoil, the time-variations of the force and moment coefficients on each airfoil are broadly similar to that of the single airfoil in that the vertical force is mainly produced in downstroke and horizontal force in upstroke, but very large differences exist due to the interaction. (3) For in-phase stroking, the major differences caused by the interaction are that the vertical force on FA in downstroke is increased and the horizontal force on FA in upstroke decreased. As a result, the magnitude of the resultant force is almost unchanged but it inclines less forward. (4) For counter stroking, the major differences are that the vertical force on AA in downstroke and horizontal force on FA in upstroke are decreased. As a result, The magnitude of the resultant force is decreased by about 20% percent but its direction is almost unchanged. (5) For 90°-phase-difference stroking, the major differences are that the vertical force on AA in downstroke and horizontal force on FA in upstroke are decreased greatly and the horizontal force on AA in upstroke increased.   As a result the magnitude of the resultant force is decreased by about 28% and it inclines more forward. (6) Among the three cases of phase angles, in-phase flapping produces the largest vertical force (also the largest resultant force); the 90°-phase-difference flapping has the largest horizontal force, although it produces the smallest resultant force.

## 1. INTRODUCTION

The flight of dragonflies is superior to that of other large insects. They are capable of fast take-off, long-time hovering and quick maneuver. Scientists have always been fascinated by their flight. Recently, the mechanism of their flight is gaining more attention due to the possible application in the newly emerging micro-air-vehicles.

Kinematic data such as the stroke amplitudes and inclinations of the stroke-planes, wing beat frequencies and phase-relations between the fore and aft wings, etc., were measured by taking high-speed pictures of tethered dragonflies (e.g. by Alexander[1]) and dragonflies in free-flight (e.g. by Norberg[2] and Wakenling and Ellington[3]). Using these data in quasi-steady analyses (not including the interaction effect between the wings), it was found that the lift coefficient required for flight were much greater than the steady-state values that measured from dragonfly wings. This suggested that unsteady wing motions and flow interactions between the fore and aft wings must play important roles in the flight of dragonflies[2-4].

Force measurement on a tethered dragonfly was conducted by Somps and Luttges[5]. It was shown that for an instant during each stroke cycle, lift force was many times larger than that measured from dragonfly wings under steady-state conditions. This clearly showed that the effects of unsteady flow and wing interaction were important. Flow visualization studies on flapping model dragonfly wings were conducted by Saharon and Luttges[6,7], and it was shown that constructive or destructive wing/flow interactions might occur, depending on the kinematic parameters of the flapping motion.

In the above works, only the total force of the fore and aft wings was measured, and furthermore, force measurements and flow visualizations were conducted in separated works. In order to further understand the dragonfly aerodynamics, it is desirable to know the aerodynamic force on each of the fore and aft wings and the flow structures during their flapping motions. In the present paper, we obtain these information by employing the method of numerically solving the Navier-Stokes equations with moving overset grids. Since the aspect-ratio of a dragonfly wing is large, as a first approximation, airfoils are used to represent the wings. Two airfoils in tandem configuration performing flapping motions in air that is still at infinity are considered in the present study. Dragonflies can alter the phase relationships between the fore and aft wings: in-phase stroking may be used to generate more aerodynamic force than the usual dragonfly mode of counter stroking (180º phase difference) and an approximate 90º phase difference may be used in escape mode where the dragonfly exhibits rapid forward and upward flight[1]. In the present study, three typical phase differences between the fore- and aft-airfoil flapping cycles, as sketched in Fig.1, are considered, namely, in-phase, 180º phase difference and 90º phase difference.

## 2. THE COMPUTATIONAL METHOD

The Navier-Stokes equations for incompressible flow are numerically solved using moving overset grids. For flow past a body in arbitrary motion, the governing equations can be cast in an inertial frame of reference using a general time-dependent coordinate transformation to account for the motion of the body.

A single-grid solver using the algorithm developed by Rogers etal.[8] was written by the present authors[9]. The algorithm is based on the method of artificial compressibility and uses a third-order flux-difference splitting technique for the convective terms and the second-order central difference for the viscous terms. Time accuracy in the numerical solutions is achieved by subiterating the equations in pseudotime for each physical time step. The single-grid solver is modified to overset grids for the present study.

With overset grids for the present study, as shown in Fig.2, each airfoil has a curvilinear grid and they lie within a background Cartesian grid. Parts of the two airfoil-grids overlap when the two airfoils move close to each other. The airfoil grids capture features such as boundary layers, separated vortices and vortex/airfoil interactions, etc. The background grid surrounds the airfoil grids and carries the solution to the far field. As a result of the oversetting of the grids, there are hole regions in the airfoil grids and the background grid. As the airfoil grids move, the holes and hole boundaries change with time. To determine hole-fringe points, the method known as domain connectivity functions by Meakin[10] is employed. Data are interpolated from one grid to another at the hole-fringe points and similarly at the outer-boundary points of the airfoil grids.   In the present study, the background grid does not move and the two airfoil-grids move in the background grid.

For far field boundary conditions, at the inflow boundary, the velocity components are specified as freestream conditions while the pressure is extrapolated from the interior; at the outflow boundary, the pressure is set equal to the free-stream static pressure and the velocity is extrapolated from the interior. On the airfoil surfaces, impermeable wall and no-slip boundary conditions were applied, and the pressure on the boundary is obtained through the normal component of the momentum equation.

The airfoil grids are generated by using a Poisson solver based on the work of Hilgenstock[11]. They are of O topology. The background Cartesian grid is generated algebraically. Some portions of the grids are shown in Fig.2.

## 3. TEST OF THE CODE

In the previous work of the present authors[9], the three dimensional version of the single-grid code was used for calculating the unsteady flows around a rotating wing and some test of the code was done there. Here the two dimensional version of the code is tested by calculating the flow around a circular cylinder, for which there exist unsteady flow measurements and computational results. In Table 1, the computed amplitudes of lift and drag coefficients and the vortex-shedding frequency (denoted by $C_L$, $C_D$ and $St$ respectively) were compared with experimental data[12] and also with the computational results of Ref.8, and they are in good agreement. A further test is done by considering the starting flow around the thin elliptical airfoil that is used in the present study, and the results will be shown later in this section.

In order to test the code for moving overset grids, flows around the airfoil was computed using both single-grid and the moving overset grids. The motion of the airfoil is similar to a flapping motion and it consists of three parts: the first translation (after initial acceleration from rest) at 35° angle of attack, about 2 chord lengths of travel; rotation; and the second translation in the opposite direction from the first with the same angle of attack. The moving overset grids consist of two grids, i.e. an airfoil grid and a background grid. During the grid motion, the background grid remains stationary as the airfoil grid translates and rotates in it. Fig.3 gives the comparison of the lift coefficients v.s. nondimensional time $\tau$, calculated by the single-grid and the moving overset grids. The large peaks in the lift coefficient are due to the fast acceleration, deceleration and rotation of the airfoil. Fig.4 compares the vorticity plots at one time instant ($\tau = 4.25$) from the two calculations. It is seen that results computed using the moving overset grids are in agreement with that of the single-grid.

At the beginning of the above motion (near $\tau = 0$ in Fig.3), the force on the airfoil should be mainly due to the acceleration of the apparent mass. Since the airfoil is an ellipse, the apparent mass force could be calculated analytically from potential flow theory and the result included in Fig.3 (the acceleration is a sine function of $\tau$, so is the apparent mass force). It is seen that between $\tau = 0$ and $\tau \approx 0.15$, the numerical results are in very good agreement with that of analytical solution. This further verified the computation code.

Grid sensitivity was considered and will be discussed together with the calculated results in the next part of the paper.

## 4. RESULTS AND DISCUSSION

The kinematics of the flapping of an airfoil is simplified as following. As sketched in Fig.1, the airfoil translates upward or downward along the stroke plane and rotates around the stroke reversals. The translational speed is assumed as a constant (denoted by $U$) during a stroke except near the end of the stroke and the beginning of the next stroke. Near the end of a stroke, the wing decelerates from speed $U$ to zero speed and then, at the beginning of the following stroke, accelerates to speed $U$. The translational speed, $u$, during the deceleration and acceleration is given by

$$u^+ = 0.5[1 + \cos(\pi(\tau - \tau_1)/(\tau_2 - \tau_1))] \qquad \tau_1 \leq \tau \leq \tau_2 \qquad (1)$$

Where $u^+ = u/U$, $\tau = tU/c$ ($c$ denotes the airfoil chord length), $\tau_1$ is the time at which the deceleration near the end of a stroke starts and $\tau_2$ the time at which the acceleration at the beginning of the next stroke finishes. During a stroke, the angle of attack on an airfoil, $\alpha$, is also assumed constant except near the stroke reversal. Around the stroke reversal, the wing rotates with angular velocity $\dot{\alpha}$. In the transition from downstroke to upstroke, $\dot{\alpha}$ is given by

$$\dot{\alpha}^+ = 0.5\dot{\alpha}_0^+ \{1 - \cos[2\pi(\tau - \tau_r)/\Delta\tau_r]\} \qquad \tau_r \leq \tau \leq \tau_r + \Delta\tau_r \qquad (2)$$

Where $\dot{\alpha}^+ = \dot{\alpha}c/U$, $\dot{\alpha}_0^+$ is a constant, $\tau_r$ is the time at which the rotation starts, $\Delta\tau_r$ is the time interval the rotation lasts. In the time interval of $\Delta\tau_r$, the wing rotates from "downstroke-$\alpha$" to "upstroke-$\alpha$", therefore when downstroke-$\alpha$, upstroke-$\alpha$ and $\Delta\tau_r$ is specified, $\dot{\alpha}_0^+$ can be determined. During the transition from upstroke to downstroke, the airfoil would rotate from upstroke-$\alpha$ to downstroke-$\alpha$ and the sign of the RHS of Eq. (2) should be reversed. The rotation-axis of each airfoil is at its 0.25 chord location. Dragonflies have small stroke angles (55º~65º) and a typical wing-section travels about 4 chord lengths in a stroke, therefore, in the present calculation the time taken for one stroke is such that the airfoil travels about 4 chord lengths in the stroke. For the cases of in-phase stroking, counter-stroking and 90º-phase difference stroking, the flapping cycle of the aft airfoil leads that of the fore airfoil by 0º, 180º and 90º respectively (Fig.1).

The airfoils are ellipses of 12 percent thickness (the radius of curvature of the airfoil leading or trailing edge is only $0.007c$). The stroke plane inclines from horizontal by $60^\circ$, downstroke-$\alpha$ is set as $60^\circ$ and upstroke-$\alpha$ $150^\circ$ (or $\alpha' = 30^\circ$, where $\alpha' = 180^\circ - \alpha$, see Fig.1). The Reynolds number, Re, defined as $Re = cU/\upsilon$ ($\upsilon$ is the kinematic viscosity), is set as 1000. The above values of inclination angle of stroke planes, $\alpha$ and Re are typical ones employed in dragonfly flight[2]. The horizontal distance between the fore and aft wings of a dragonfly is about zero at the wing root and about $0.5\,c$ at the wing-tip. Therefore in this study, we used the average value, $0.25c$.

The coefficients of vertical and horizontal forces and pitching moment (taken with respect to 0.25 chord location) are denoted by $C_y$, $C_h$ and $C_m$ respectively(the force and moment coefficients are obtained by dividing the force and moment with $\rho U^2 c/2$ and $\rho U^2 c^2/2$ respectively, where $\rho$ is the fluid density). For convenience, the horizontal force is taken as positive when it is in the negative x-direction. As seen in Figs.1 and 2, both the lift (normal to the translational velocity) and drag (opposite to the translational velocity) of the airfoil contribute to the vertical force and to the horizontal force. In the calculation, the airfoils start the flapping motion in still air and the calculation is stopped when periodicity in the aerodynamic force and flow structures is approximately reached. In order to reveal the interaction effect between the two airfoils, flow for a single airfoil of the same geometry and in the same motion is also calculated.

**4.1 Single Airfoil**

The results for the single airfoil (henceforth called SA), can be used to compare with that of the two airfoils to show the interaction effect. Moreover, understanding the force generation mechanism of SA provides background for studying the more complex case of the fore and aft airfoils.

Figure 5 gives the force and moment coefficients of SA vs. $\tau$ in one cycle. The motion of the airfoil, i.e. $u^+$ and $\dot\alpha^+$ as functions of $\tau$, is also shown in Fig. 5. Fig. 6 gives the vorticity and streamline plots at various $\tau$. In the vorlicity plots of this paper, solid lines denote positive vorticity and dotted lines negative vorticity. In Fig. 5, results by two overset-grids are given to show the grid sensitivity. In both overset-grids, the airfoil-grid extends about $0.9\,c$ from the airfoil surface; the Cartesian background grid has grid points concentrated in the near field of the stroke planes (the density here is approximately the same as that near the outer boundary of the airfoil grid), and it extends to about 40 chord lengths from the edge of the stroke plane. For overset-grids 1, the airfoil grid is the size of $55\times153$ (in normal direction and around the airfoil respectively) and the background Cartesian grid is the size of $233\times114$. For overset-grids 2, the airfoil grid is the size of $76\times229$ and the background Cartesian grid is the size of $291\times132$. It is seen that there is only a small difference between the results of the two grids. Calculation was also conducted using larger computational domain. The domain is enlarged by adding more grid points to the background Cartesian grid of overset-grids 1. The calculated results showed that there was no need to put the outer boundary farther than 40 chord lengths from the stroke planes. From the above discussion, it was concluded that overset-grids 1 is proper for the single-airfoil calculation. For the overset grids for two airfoils, each of the airfoil grids is the same as that of overset-grids 1 and the background grid is constructed in a similar way to that of overset-grids 1. The effect of time step value was considered and it was found that a numerical solution effectively independent of time step was achieved if $\Delta\tau\le0.01$. Therefore, this time step value was specified in all the present calculations.

As seen in Fig.5, in the beginning part of the downstroke ($\tau$=26.4~27.45), there are a large peak in $C_Y$ and a moderate negative one in $C_H$. In this part of the stroke, the airfoil accelerates downwards along the stroke plane and continues the anticlockwise rotation that started near the end of the upstroke. It was shown by Hamdani and Sun[13] that for an airfoil in fast translational acceleration at large angle of attack, large aerodynamic force, which is approximately normal to the airfoil chord, could be produced. This may explain the above $C_Y$ and $C_H$ behaviors. In the next part of the downstorke ($\tau$=27.45~29.75), the airfoil translates downwards with constant speed at $\alpha=60^\circ$ and its chord is horizontal (Fig.6(a)~(c)). In this part of the stroke, $C_Y$ keeps a large value, about 2.7. This is expected because as shown in the previous work[13], for an airfoil at large α moving at constant speed after a fast start, large aerodynamic force can be maintained for some time due to the stall delay. As seen in Fig.6(a) to (c), the dynamic stall vortex has not shed yet. $C_H$ is very small in this part of the motion and the total force is approximately normal to the airfoil. From the vorticity plot at $\tau$=27.45, Fig.6(a) (just after the acceleration and rotation finish), it is seen that a vortical structure (a strong vortex layer on the lower surface of the airfoil and a pair of vortices near the trailing edge of the airfoil) has been formed. The induce velocity of this vortical structure makes the effective angle of attack to become about $35^\circ$(as seen from the streamline plot in Fig. 6(a)), which is not small. If a smaller geometrical angle of attack had been used, the effective angle of attack would be too small. This may be the reason for dragonflies to employ very large angle of attack in the downstrokes. In the last part of the downstroke ($\tau$ =29.75~30.8), $C_Y$ first increases rapidly and then decreases sharply, with the peak at τ ≈ 30.2. From $\tau$ =29.75 to $\tau$ =30.1, the airfoil rotates clockwise (pitching-up with respect to the direction of translational motion) while still translating with constant speed. As shown in Ref. 13, this motion can produce fast increases in lift and drag forces. This explains the rapid increases of $C_Y$. Between $\tau$ =30.1 and $\tau$ =30.8, the airfoil is in fast deceleration, which would cause sharp decrease in lift and drag forces of an airfoil[13], resulting in the rapid decreases of $C_Y$ in this period.

In the beginning part of the upstroke ($\tau$ =30.8~31.85), there are a negative peak in $C_Y$, and a positive peak and then a negative one in $C_H$. In this period of time, the airfoil accelerates upwards along the stroke plane and at the same time rotates clockwise from $\alpha' = 75°$ at τ =30.8 to $\alpha' = 30°$ at $\tau$ =31.85 (the airfoil chord becomes vertical at $\tau$ = 31.85). If the airfoil only accelerates upward along the stroke plane, one would expect large positive value for $C_H$ and negative values for $C_Y$. But the clockwise-rotation of the airfoil (with rotation-axis near the airfoil leading edge) has opposite effect from that of the acceleration, and furthermore, the orientation of the airfoil changes, resulting in the positive and negative peaks in $C_H$ and the negative peak in $C_Y$. In the next part of the upstroke ($\tau$ =31.85~34.15), the airfoil translates upwards along the stroke plane with constant speed at $\alpha' = 30°$ and its chord is vertical (Fig.6(d) to (f)). From the previous discussion, one would expect that in this period, $C_H$ (which is now normal to the airfoil chord) is large and almost does not vary with $\tau$. But as seen in Fig. 5, $C_H$ is small at $\tau$ =31.85 and increases with $\tau$, reaching large values in later time of this period. The reason for this is that the induced velocity ("downwash"), which is due to the vortical structure formed after the acceleration and rotation, has made the effective angle of attack almost zero (as seen from the streamline plot in Fig.6(d), the incoming streamlines are almost aline with the airfoil chord). As $\tau$ increasing, the above mentioned vortical structure move away from the airfoil (Fig.6(e) and (f)) and the effective angle of attack increases. In the last part of the upstroke ($\tau$ =38.55~39.6), large peaks in $C_H$ and $C_Y$ appear (Fig.5), which are caused, similar to the case of downstroke, by the fast rotation and the following deceleration of the airfoil.

The above discussion is summarized as following. For an airfoil in the flapping motion considered in the paper, positive vertical force is mainly produced in the downstroke and positive horizontal force in the upstroke. The time-averages of $C_y$ over the downstroke (denoted by $\overline{C}_{y,d}$) and the time average of $C_H$ over the upstroke (denoted by $\overline{C}_{H,u}$) are given in Table 2. $\overline{C}_{y,d}$ is 2.74. This is much larger than the maximum lift coefficient measured from dragonfly wings, which are about 1. $\overline{C}_{H,u}$ is 1.97 and is also large. The main reasons for the large force coefficients are: the acceleration at the beginning of the stroke, the delayed stall, and the pitching-up motion near the end of the stroke. The vortices formed after the stroke reversal cause large induced velocity, which greatly reduces the effective angle of attack of the airfoil in the early part of the following stroke. This may be the reason for dragonflies to employ very large geometrical angle of attack during the downstroke.

The vertical and horizontal force coefficients averaged over one complete flapping cycle for SA, denoted by $\overline{C}_y$ and $\overline{C}_H$ respectively, are given in Table 3. It is seen that for SA, $\overline{C}_y$ is 1.22, about one half of its $\overline{C}_{y,d}$ (since $C_y$ is mainly produced during the downstroke); $\overline{C}_H$ is 0.76, also about one half of $\overline{C}_{H,u}$.

## 4.2 Fore and aft airfoils

Henceforth the fore and aft airfoils are called FA and AA respectively. As will be seen shortly, the time-variations of the force and moment coefficients on FA and AA are broadly similar to that of SA, but large quantitative and in some parts of the flapping cycle qualitative differences exist. The aerodynamic forces on FA or AA can be considered as contributed by two factors, the unsteady motion of the airfoil and the aerodynamic interaction between the two airfoils. The general time-variations of the force and moment coefficients of FA or AA, which is similar to that of SA, is due to the unsteady motion of the airfoil. The above differences between the force and moment coefficients of FA (or AA) and that of SA are due to the aerodynamic interaction. Since the general time-variations of force and moment coefficients of FA and AA are similar to that of SA and the later were investigated in detail in last section, we will mainly study the aerodynamic interaction between the two airfoils in this section.

*0° phase difference (in-phase stroking)*

Figure 7 gives the force and moment coefficients v.s. time in one flapping cycle (the results for SA is included for comparison). Figures 8, 9 and 10 show the flow structures.

In the downstroke, as seen in Fig. 7, $C_H$ of FA and AA are small (except at the beginning and the end phases of the stroke) and relatively close to that of SA. In some parts of the downstroke, $C_y$ of FA and AA are different from that of SA. In the beginning part of the downstroke ($\tau$ =26.4~27.45), $C_Y$ of FA is a little smaller than that of SA but $C_Y$ of AA has a much larger peak. In this period of time, FA and AA accelerate downward along the stroke planes and at the same time rotate anticlockwise from $\alpha = 105°$ to $\alpha = 60°$. FA moves in front of AA (see Fig.1(a)) and the rotation of FA makes its rear part to move towards the leading edge of AA. This could result in a larger "incoming velocity" seen by AA. Figure 8 (a) and (b) gives the relative-velocity vectors seen by FA and by AA, respectively (since FA and AA have relative motion in this part of the downstroke, their relative velocity vectors need to be plotted on separated plots). It is seen that a much larger velocity going around the leading edge of AA than that of FA, which may explain the very large peak in $C_Y$ of AA in this period. In the next part of the downstroke ($\tau$ =27.45~29.75), at first $C_Y$ of both FA and AA are larger than that of SA, but as $\tau$ increasing, $C_Y$ of AA decreases to become smaller than that of SA. In this period, FA and AA move downward at constant-speed with their chords in the same horizontal line (see Fig.1(a)). As $\tau$ increasing, the trailing edge vortex of FA (which has positive vorticity) moves to the neighborhood of the upper surface of AA and interacts with its leading edge vortex (which has negative vorticity), see Fig.9(b). This vortex-interaction will reduce the circulation around AA. Moreover, AA acts like an extension of FA. As a result, the flow develops into one like that of a slotted two-element airfoil (with FA and AA as its fore and rear parts respectively). The surface pressure distribution at $\tau = 29$ (given in Fig.11) also shows this point. The above discussion explains why, in this period, $C_Y$ of FA is larger but $C_Y$ of AA is smaller than that of SA. In the last part of the downstroke ($\tau$ =29.75~30.8), similar to the case of SA, $C_Y$ of both FA and AA have large peaks, due to the fast pitching-up rotation and the following fast deceleration of the airfoils. As shown Table 2, the time average of $C_Y$ over the downstroke for FA is 3.49, about 25 percent larger than that of SA and for AA is 2.57, a little smaller than that of SA.

In the upstroke, $C_Y$ of both FA and AA are small and very close to that of SA.  In the beginning part of the stroke ($\tau$ =30.08~31.13), $C_H$ of FA is larger and $C_H$ of AA smaller than that of SA. This may be due to the relative motion of the two airfoils which are in clockwise rotation and upward acceleration. In the rest part of the stroke (constant-speed translation and the following anticlockwise rotation), $C_H$ of AA is close to that of SA and $C_H$ of FA is noticeably smaller than that of SA. In the constant-speed translation, FA and AA moves parallely upward along the stroke planes and their chords are vertical, as seen in Fig.9 (d), (e) and (f). It should be noted that AA moves in front of FA in the upstroke (see Fig.1(a)). Due to the induction effect of the positive vorticities at the upper surface of AA (see Fig. 9(e) and (f)), there will be "downwash velocity" in front of FA, making its effective angle of attack smaller. This can be seen from the streamline plots shown in Fig.10(e) and (f) . It is seen that the effective angle of attack of FA is smaller than that of AA. This explains why $C_H$ of FA is smaller during this phase of the upstroke. The same explanation also applies in the following anticlockwise rotation (pitching-up with respective to the direction of translational motion). As shown in Table 2, $\overline{C}_{H,u}$ for both FA and AA are smaller than that of SA, but $\overline{C}_{H,u}$ for FA is much smaller.

For the cases of two airfoils, $\overline{C}_y$ and $\overline{C}_H$ denote, respectively, the vertical and horizontal force-coefficients averaged over one complete flapping cycle and also averaged between the fore and aft airfoils. As seen in Table 3, for the present case (in-phase stroking), $\overline{C}_y$ is larger than that of SA. The main reason for this is that, as seen from above discussion, $\overline{C}_y$ on FA in the downstroke is increased greatly by the interaction affect. $\overline{C}_H$ is smaller than that of SA and the main reason for it is that $C_H$ on FA in the upstroke is greatly decreased by the interaction effect.

*180° phase difference* (*counter stroking*)

Figure 12 gives the force and moment coefficients v.s. time in one flapping cycle. It is seen that in the first half of the cycle (AA in downstroke and FA in upstroke), $C_y$ of AA is smaller than that of SA (Fig.5), also smaller than that of AA in in-phase stroking (Fig.7). $\overline{C}_{y,d}$ of AA is only 1.85. $C_H$ of FA is smaller than that of SA except at the beginning of the stroke; $C_{H,u}$ is only 1.34. Figure 13 shows the vorticity plots. In this and some of the following figures, the vortices shed by FA are marked by Fn or F and the vortices shed by AA are marked by An or A, where n is an integer. It is seen that in this half cycle, Fig.13(a), (b) and (c), vortices carried and shed by FA may produce downwash velocity around AA and vice versa. This can be seen in the streamline plots in Fig.14(a),(b) and (c), the effective angle of attack of AA is smaller than that of SA in its downstroke (Fig.6(a), (b) and (c)) and the effective angle of attack of FA is also smaller than that of SA in its upstroke (Fig.6(d), (e) and (f)), this may explain the above force behaviors.

In the second half of the cycle (AA in upstroke and FA in downstroke), as seen in Fig.12, $C_y$ of FA is small in the early part of the stroke but becomes large in the later part of the stroke ($\overline{C}_{y,d}$ of FA is 2.61); $C_H$ of AA is negative in a large part of the stroke and then sharply increases to large value ($\overline{C}_{H,u}$ of AA is 1.67). From the vorticity plots, it is seen that in the early part of the downstroke of FA, Fig.13(d), compared with the case of SA (Fig.6(a)), there is one more pair of vortex near the trailing edge of FA. This vortex pair, marked by A1 and A2, is left by AA in its last stroke, and it can induce a wind that blows leftwards over FA. This wind may decrease the downstream motion of the starting vortex of FA (marked by F5), resulting in the small $C_y$ of FA in this part of the stroke. In the early part of the upstroke of AA, Fig.13(d), compared with the case of SA (Fig.6(d)), there is one more vortex pair (marked by A4 and F4) and one more vortex (marked by F3) around AA, which are left by FA and AA in their last strokes. It is clear that these vortices will produce a "down wash" velocity in front of AA. As seen in the streamline plot for AA at this instance, Fig.14(d), the effective angle of attack of AA is negative. As τ increasing, AA moves upwards, away from these vortices, but it becomes close to FA which is moving downwards. The strong vortices carried by FA, marked by F5 and F7 (Fig.13(d) and (e)), may also produce "downwash" effect on AA. From Fig.14(e), it is seen that even at τ =50.6, a little beyond the middle of the upstroke of AA, the effective angle of attack of AA is still negative. This may explain why $C_H$ of AA is negative in this part of the stroke. After τ =50.6, AA and FA have passed each other. It is interesting to see that AA "takes" the starting vortex of FA, F5 (which has positive vorticity), and carries the vortex on its upper surface, Fig.13(e) and (f). This vortex may increase the circulation of AA greatly, which may explain the sharp increase of $C_H$ on AA in this part of the stroke.

For the present case of counter stroking, as seen in Table 3, $\overline{C}_y$ and $\overline{C}_H$ are both smaller than of SA. From the above analysis and the data in Table 2, it is seen that the main reason for this is that, during the first half of the flapping cycle (AA in downstroke, FA in upstroke), $C_y$ on AA and $C_H$ on FA are greatly decreased by the interaction effect.

*90° phase difference*

Figure 15 gives the force and moment coefficients v. s. time in one flapping cycle and Figs. 16 and 17 give the flow structures.

In the downstroke of AA (the second half of the upstroke and the first half of the downstroke of FA), τ =44~48.4, as seen from Fig.15, $C_y$ of AA is much smaller than that of SA in its downstroke and $C_H$ of FA is also smaller than that of SA in the second half of its upstroke. From the vorticity plots in Fig.16(a), (b) and (c), it is seen that similar to the case of counter stroking, vortices carried and shed by FA may produce induced velocity around AA and vice versa; moreover, in the present case, FA and AA are more closer to each other. The interaction effect may become more severe. As seen in the streamline plots, Fig.17 (a), (b) and (c), the effective angle of attack of AA is much smaller than that of SA (Fig.6 (a), (b) and (c)). As seen from Fig.17(a), the effective angle of attack of FA is also smaller than that of SA at this position (Fig.6(e)). This may approximately explain the above force behaviors. As seen in Table 2, $\overline{C}_{y,d}$ of AA is only 1.5 and $\overline{C}_{H,u}$ of FA is only 0.72. They are the smallest among the cases considered.

In the upstroke of AA  (the second half of the downstroke and first half of the upstroke of FA), $\tau$ =48.4~52.8, as seen in Fig.15, the most noticeable behavior of the force coefficients is that starting from $\tau$ =49.5, $C_H$ of AA becomes larger than that of SA in its upstroke. From the vorticity plots, Fig.16(e) and (f), it is seen that in this part of the stroke, there are some vortices (F5 and A1) in front of AA. They are in such a position that they may produce a wind that blows in the direction opposite to the motion of AA. This wind could increase the incoming-flow speed of AA and thus its $C_H$. As seen in Table 2, $\overline{C}_{H,u}$ of AA is 2, the largest in all the cases considered.

For the present case, $\overline{C}_y$ is only 0.78 (see Table 3), the smallest in all the cases. As analyzed above, the reason for this is that due to the interaction effect, FA and AA have small $C_y$ in their downstrokes. $\overline{C}_H$ is 0.68, relatively large compared with other cases, which is because AA has a relatively large $C_H$ in the upstroke due to the interaction affect.

Finally we look at the resultant of $\overline{C}_y$ and $\overline{C}_H$. Let $C_R$ denote the magnitude of the resultant and $\beta$ the angle between the resultant and the vertical (positive when the resultant incline forward), and they are also shown in Table 3. Comparing the results for the three cases of two airfoils with that of SA, the effect on $C_R$ and $\beta$ due to the interaction can be seen. For the case of in-phase stroking, the interaction almost does not change the magnitude of the resultant and only makes it to incline less forward; for the case of counter stroking, the interaction decreases the magnitude of the resultant by about 20% and almost does not change its direction; for the case of 90˚-phase-difference stroking, the interaction decreases the magnitude of the resultant by about 28% and makes it to incline more forward. Among the three cases of two airfoils in flapping motion, the in-phase stroking produces the largest vertical force (also largest resultant force) and the 90˚-phase-difference stroking has the largest horizontal force, although it produces the smallest the resultant force.

## 5. CONCLUSIONS

For the flapping motion considered in the paper, the following has been shown.

(1) In the case of no interaction (single airfoil), the time average of the vertical force coefficient over the downstroke is 2.74, which is about 3 times as large as the maximum steady-state lift coefficient of a dragonfly wing; the time average of the horizontal force coefficient is 1.97, which is also large. The reasons for the large force coefficients are the acceleration at the beginning of a stroke, the delayed stall and the "pitching-up" motion near the end of the stroke.

(2) In the cases of two-airfoil, the time-variations of the force and moment coefficients on each airfoil are broadly similar to that of the single airfoil in that the vertical force is mainly produced in downstroke and horizontal force in upstroke, but very large differences exist due to the interaction.

(3) For in-phase stroking, the major differences caused by the interaction are that the vertical force on FA in downstroke is increased and the horizontal force on FA in upstroke decreased. As a result, the magnitude of the resultant force is almost unchanged but it inclines less forward.

(4) For counter stroking, the major differences are that the vertical force on AA in downstroke and horizontal force on FA in upstroke are decreased. As a result, The magnitude of the resultant force is decreased by about 20% percent but its direction is almost unchanged.

(5) For 90˚-phase-difference stroking, the major differences are that the vertical force

on AA in downstroke and horizontal force on FA in upstroke are decreased greatly and the horizontal force on AA in upstroke increased. As a result the magnitude of the resultant force is decreased by about 28% and it inclines more forward.

(6) Among the three cases of phase angles, in-phase flapping produces the largest vertical force (also the largest resultant force); the 90°-phase-difference flapping has the largest horizontal force, although it produces the smallest resultant force.

Table 1    Lift and drag coefficients and Strouhal numbers for circular cylinder flow at Reynolds number 200

|  | $C_D$ | $C_L$ | $S_t$ |
|---|---|---|---|
| Present | $1.225 \pm 0.035$ | $\pm 0.62$ | 0.19 |
| Kovasznay[12] | — — — | — — — | 0.19 |
| Rogers[8] | $1.23 \pm 0.05$ | $\pm 0.65$ | 0.185 |

Table 2 Values of $C_y$ averaged over downstroke and $C_H$ over upstroke.

| phase difference | $\overline{C}_{y,d}$ | $\overline{C}_{H,u}$ |
|---|---|---|
|  | 2.74 (SA) | 1.97 (SA) |
| $0°$ | 3.49 (FA) | 1.21 (FA) |
|  | 2.57 (AA) | 1.56 (AA) |
| $180°$ | 2.61 (FA) | 1.34 (FA) |
|  | 1.85 (AA) | 1.67 (AA) |
| $90°$ | 2.03 (FA) | 0.72 (FA) |
|  | 1.50 (AA) | 2.00 (AA) |

Table 3 The total vertical and horizontal force coefficients and the resultant force coefficients.

| phase difference | $\overline{C}_y$ | $\overline{C}_H$ | $C_R$ | $\beta$ |
|---|---|---|---|---|
| (SA) | 1.22 | 0.76 | 1.44 | $32°$ |
| $0°$ | 1.34 | 0.46 | 1.42 | $18°$ |
| $180°$ | 0.95 | 0.62 | 1.12 | $33°$ |
| $90°$ | 0.78 | 0.68 | 1.03 | $41°$ |

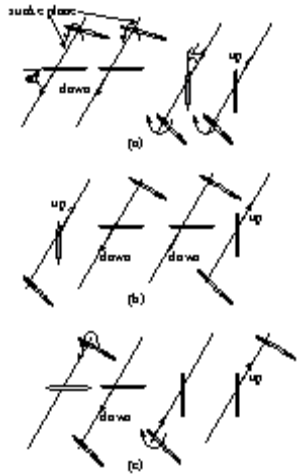

Fig.1 A sketch of the flapping motions. (a) in-phase, (b) 180°phase difference, (c) 90°phase difference.

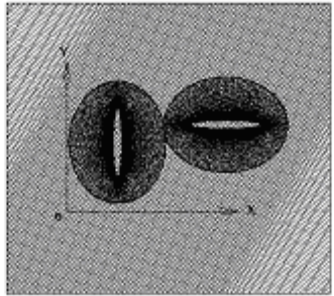

Fig.2 Some portions of the moving overset grids.

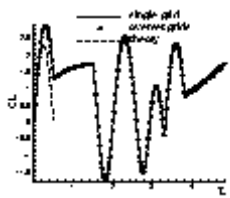

Fig.3 Comparison between the lift coefficients of a moving airfoil calculated using single grid and moving overset grids. CL contributed by apparent mass force, obtained from potential flow theory, is included (the theory is applicable only near $\tau = 0$).

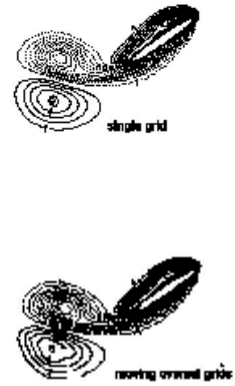

Fig.4 Comparison between the vorticity contours of a moving airfoil at $\tau = 4.25$ calculated using single grid and moving overset grid.

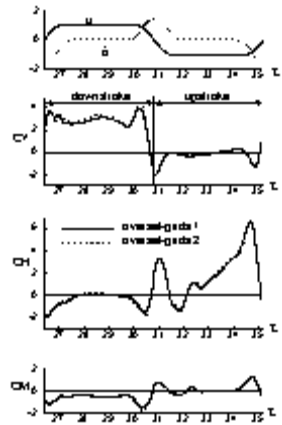

Fig.5 Force and moment coefficients vs. $\tau$, single airfoil.

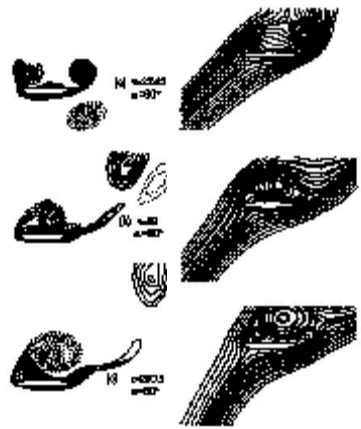

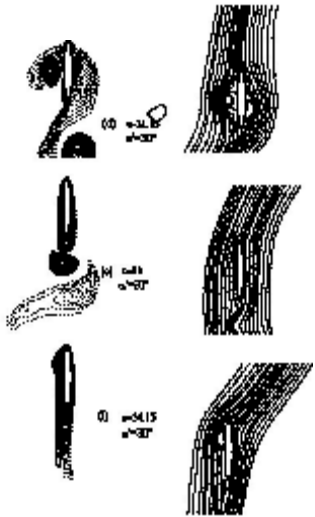

Fig.6 Vorticity and streamline plots, single airfoil

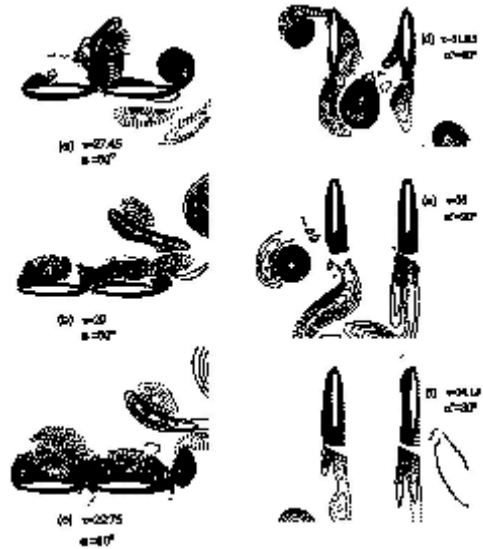

Fig.9 Vorticity plots, fore and aft airfoils, in-phase.

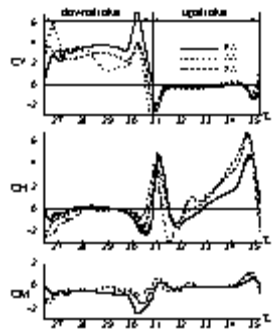

Fig.7 Force and moment coefficients vs. $\tau$ , fore and aft airfoils, in-phase.

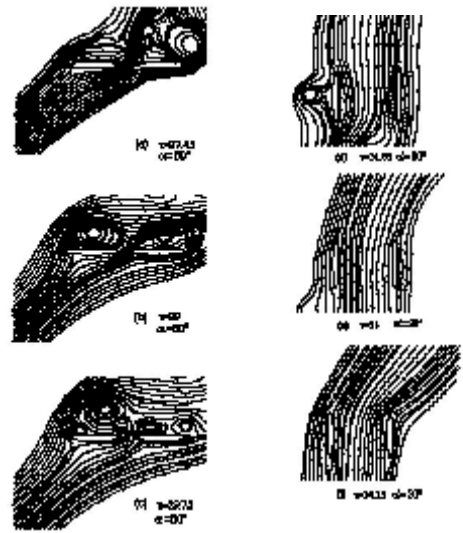

Fig.10 Streamline plots in the upstroke, fore and aft airfoils, in-phase.

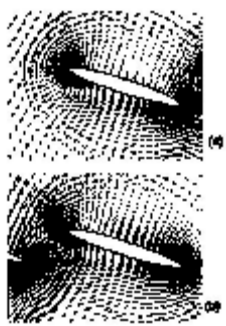

Fig.8 Velocity vector plots at $\tau = 26.75$ , in-phase.

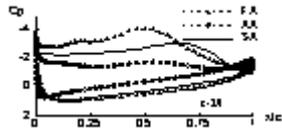

Fig.11 Surface pressure distribution
at $\tau = 29$, in-phase

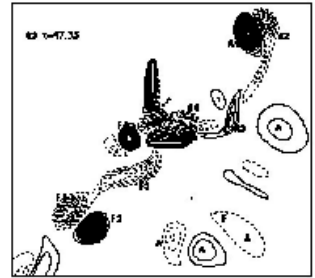

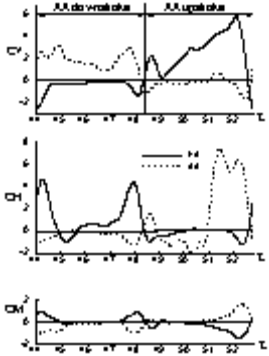

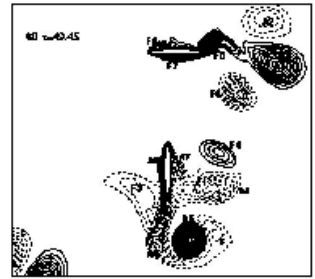

Fig.12 Force and moment coefficients vs. $\tau$.
Fore and aft airfoils,180°phase difference

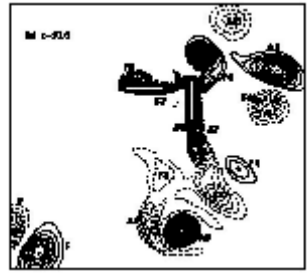

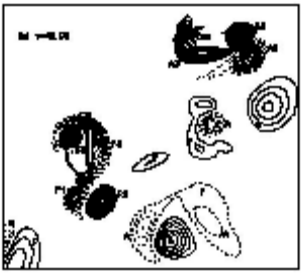

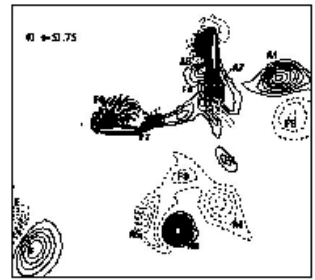

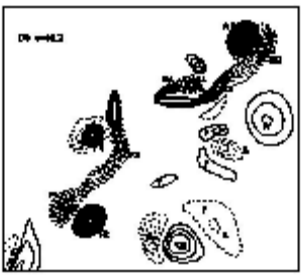

Fig.13 Vorticity plots at various times. Fore and aft
airfoils, 180° phase difference

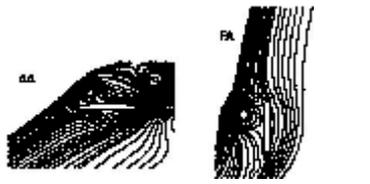

(a) $\tau = 45.05$

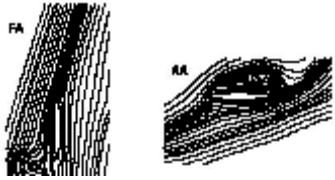

(b) $\tau = 46.2$

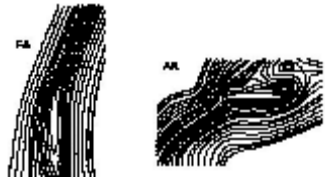

(c) $\tau = 47.35$

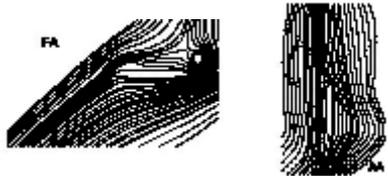

(d) $\tau = 49.45$

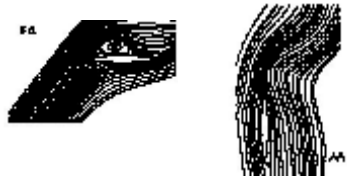

(e) $\tau = 50.6$

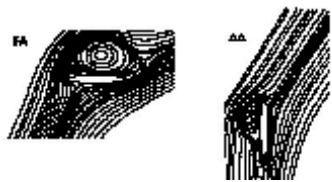

(f) $\tau = 51.75$

Fig.14 Streamline plots at various times. Fore and aft

airfoils, 180°phase difference

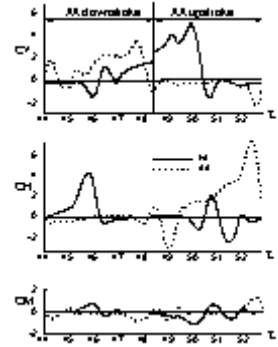

Fig.15 Force and moment coefficients vs. $\tau$.

Fore and aft airfoils, $90^\circ$ phase difference

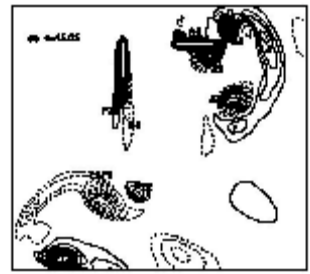

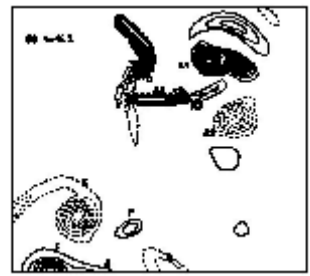

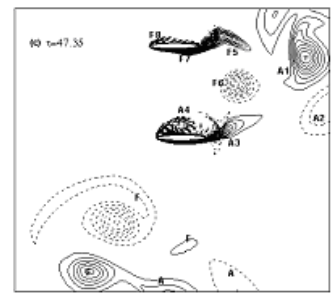

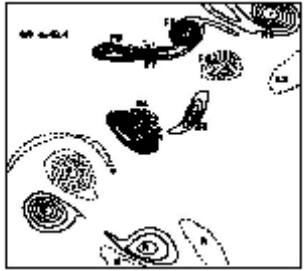

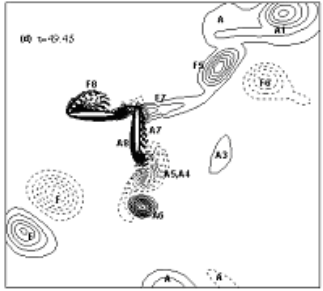

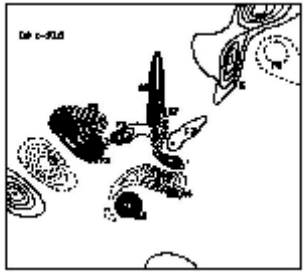

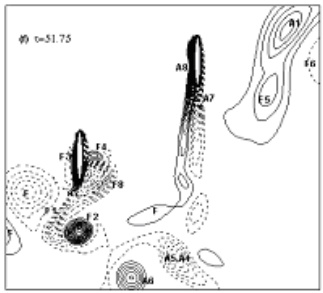

Fig.16 Vorticity plots at various time.
Fore and aft airfoils, $90°$ phase difference.

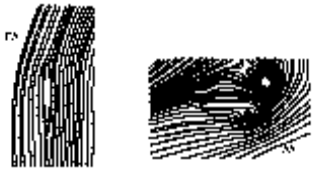

(a) $\tau = 45.05$

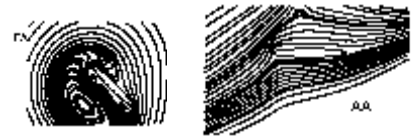

(b) $\tau = 46.2$

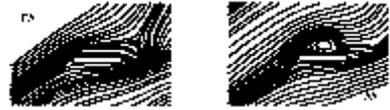

(c) $\tau = 47.35$

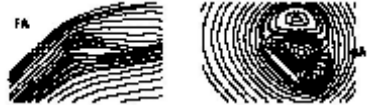

(d) $\tau = 48.4$

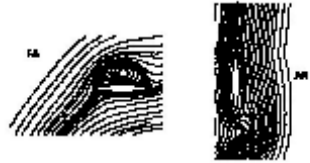

(e) $\tau = 49.45$

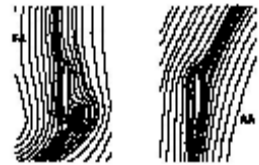

(f) $\tau = 51.75$

Fig.17 Streamline plots at various times.
Fore and aft airfoils, $90°$ phase difference.